\documentclass[final,5p,times,twocolumn]{elsarticle}
\usepackage{euscript,amssymb,amsmath}

\usepackage{graphicx}
\usepackage{epsfig}
\usepackage{color}

\newcommand{\be}[1]{\begin{equation}\label{#1}}
\newcommand{\ee}{\end{equation}}
\newcommand{\ba}[1]{\begin{eqnarray}\label{#1}}
\newcommand{\ea}{\end{eqnarray}}
\newcommand{\rf}[1]{(\ref{#1})}
\newcommand{\nn}{\nonumber}

\newcommand{\de}{\partial}

\newcommand{\hg}{\hat g}
\newcommand{\hR}{\hat R}

\newcommand{\La}{\Lambda}

\newcommand{\ve}{\varepsilon}
\newcommand{\om}{\omega}
\newcommand{\Om}{\Omega}

\begin{document}

\begin{frontmatter}

\title{Black branes and black strings in the astrophysical and cosmological context}

\author[a]{\"{O}zg\"{u}r Akarsu}
\ead{akarsuo@itu.edu.tr}

\author[b]{Alexey Chopovsky}
\ead{a.chopovsky@yandex.ru}


\author[a,b]{Alexander Zhuk}
\ead{ai.zhuk2@gmail.com}

\address[a]{Department of Physics, \.{I}stanbul Technical University, 34469 Maslak, \.{I}stanbul, Turkey\\}

\address[b]{Astronomical Observatory, Odessa National University, Dvoryanskaya st. 2, Odessa 65082, Ukraine}


%
\begin{abstract} We consider Kaluza-Klein models where internal
spaces are compact flat or curved Einstein spaces. This background is perturbed by a compact gravitating body with the dust-like equation of state (EoS) in the
external/our space and an arbitrary EoS parameter $\Om$ in the internal space. Without imposing any restrictions on the form of the perturbed metric and the
distribution of the perturbed energy densities, we perform the general analysis of the Einstein and conservation equations in the weak-field limit. All
conclusions follow from this analysis. For example, we demonstrate that the perturbed model is static and perturbed metric preserves the block-diagonal form. In a
particular case $\Om=-1/2$, the found solution corresponds to the weak-field limit of the black strings/branes. The black strings/branes are compact gravitating
objects which have the topology (four-dimensional Schwarzschild spacetime)$\times$ ($d$-dimensional internal space) with $d\geq 1$. We present the arguments in
favour of these objects. First,  they satisfy the gravitational tests for the parameterized post-Newtonian parameter $\gamma$ at the same level of accuracy as
General Relativity. Second, they are preferable from the thermodynamical point of view. Third, averaging over the Universe, they do not destroy the stabilization
of the internal space.  These are the astrophysical and cosmological aspects of the black strings/branes.
\end{abstract}


\end{frontmatter}

\section{Introduction}

\setcounter{equation}{0}

Modern physical science faces a number of fundamental problems that it cannot yet solve in the framework of standard theories. For example, the numerous
observations show that dark energy and dark matter dominate in our Universe, but their nature remains unclear for us. Numerous attempts have been made to solve
these problems. One of possible directions is to modify the theory of gravity assuming, e.g., the multidimensionality of our world. These are the so-called
Kaluza-Klein (KK) models. Obviously, on the one hand, the modified theories must respond to modern challenges, and, on the other hand, do not contradict the
observed (and well confirmed) picture of the world. In the present paper we will not concern the problems of dark energy and dark matter in KK models (some
aspects of it were discussed, e.g., in \cite{Gunther:2003yx}) but we will concentrate on the problem of viability of the KK models. We intend to check whether KK models
adequately describe astrophysical objects such as our Sun and, additionally, to show that they can correspond to the conventional cosmological model. In a number
of our previous papers (see, e.g., \cite{Eingorn:2010wi,Chopovsky:2011hp,Eingorn:2012ef,Chopovsky:2012qb,Eingorn:2012ip,Chopovsky:2014yxa}) we have already discussed the astrophysical aspects of KK models. However, in
these papers we have imposed additional conditions on considered models, e.g., the preservation of the block-diagonal form of the perturbed metric as well as the
assumption on the uniform smearing of the gravitating body over the internal space. In the present paper we do not impose these conditions from the very beginning
and perform the most general analysis for the considered model. All conclusions on the form of perturbed metric as well as on the energy density distribution of
the gravitating body follow from the analysis of the Einstein equation, the gauge condition and the conservation equation.

We present arguments that black strings/branes are preferable from the astrophysical and cosmological points of view. The black strings/branes are compact
gravitating objects which have topology (four-dimensional Schwarzschild spacetime) $\times$ ($d$-dimensional internal space) with $d\geq 1$ \cite{Traschen:2001pb,Townsend:2001rg,Harmark:2004ch,Kastor:2006ti},
$d=1$ and $d\geq 2$ in the cases of the black strings and black branes, correspondingly. They have the dust-like equation of state (EoS) in the external/our space
and  EoS $p=-1/2\epsilon$ (i.e. tension) in the internal space. The exact form for the metric coefficients was found, e.g., in \cite{Eingorn:2010wi,Eingorn:2011vu} (flat internal
space) and in \cite{Chopovsky:2012qb} (two-sphere internal space). In the present paper we obtain the metric coefficients in the weak-field limit in the case of the Einstein
internal space (flat or curved). The advantages of the black strings/branes are the following. First,  they satisfy the gravitational tests for the parameterized
post-Newtonian (PPN) parameter $\gamma$ at the same level of accuracy as General Relativity. Second, they are preferable from the thermodynamical point of view
\cite{Eingorn:2012ip}. Third, averaging over the Universe, they do not destroy the stabilization of the internal space. It is worth noting that in the case of curved
internal space the agreement with the experimental restrictions on the PPN parameter $\gamma$ can be also achieved for the EoS parameter different from $-1/2$
\cite{Chopovsky:2014yxa}. However, all three advantages take place simultaneously only for $-1/2$.   These are the astrophysical and cosmological aspects of the black
strings/branes.

The paper is structured as follows. In section 2, we describe the background model. In section 3, we consider the perturbed model and investigate the perturbed
conservation equation. Section 4 is devoted to the analysis of the perturbed Einstein equation.  Here, we present the final system of equations for the perturbed
metric coefficients and perturbed energy densities. In section 5, we analyze this system and present arguments in favour of the black strings/branes. The main
results are summarized in the concluding section 6.



\section{Background model}

As we have already mentioned in the Introduction, we intend to investigate our gravitational model in the astrophysical context. In other words, we want to check
whether our model adequately describes the gravitational fields of astrophysical objects. Obviously, the most studied astrophysical object is our Sun. Therefore,
it makes sense to check the compatibility of the well known tests in the Solar system with the predictions of our model. We have in mind tests of the
parameterized post-Newtonian (PPN) parameter $\gamma$. It follows from these experiments that $\gamma$ is extremely close to unity. For example, the Shapiro
time-delay experiment using the Cassini spacecraft gives \cite{Jain:2010ka}: $\gamma -1 = \left(2.1 \pm 2.3\right)\times 10^{-5}$. In General Relativity (GR), the
parameter $\gamma$ is exactly equal to unity which is in very good agreement with these observations \cite{Landau,Will}. Since in the Solar System the
gravitational field is weak, to calculate $\gamma$, it is usually assumed that the background 4-D spacetime geometry is flat \cite{Landau}. We also assume that
4-D part of the background metric in our KK model is flat. However, we do not know the topology and geometry of the internal space. It can be either flat or curved. Therefore,
we suppose that the internal space is an arbitrary Einstein space. This is a fairly general assumption. Hence, our background metric has the following
block-diagonal form:
\begin{equation}
\label{1.1}
\begin{aligned}
\hat g_{MN}dX^M\otimes dX^N=\hat g_{\mu\nu}dx^\mu\otimes dx^\nu+\hat g_{mn}dy^m\otimes dy^n\, ,\\
M,N=0,1,\ldots,4+d; \quad \mu,\nu=0,1,2,3;\\
m,n=4,5,\ldots,4+d\, ,
\end{aligned}
\end{equation}
where $\hat g_{\mu\nu}\equiv \eta_{\mu\nu}=\mathrm{diag (1,-1,-1,-1)}$ is the Minkowski spacetime metric. Hereafter, the hats denote
background values.
The internal space is a compact Einstein one:
\be{1.2}
\hR_{mn}[\hg^{(d)}]=\mathcal{C}\hg_{mn},\quad\hR^m_m[\hg^{(d)}]=\hR^{(d)}=\mathcal{C} d, \quad
\mathcal{C}\equiv{\rm const}. \ee
According to our sign convention, $\mathcal{C}<0$ ($\mathcal{C} >0$) corresponds to the compact Einstein space with positive (negative) curvature. For example, in
a particular case of the $d$-dimensional sphere of radius $a$ we have $\mathcal{C} = -(d-1)/a^2$.

Obviously, the metric \rf{1.1} should satisfy the Einstein equations. This means that to have a curved internal space, we need to introduce a background matter.
The structure of the energy-momentum of this matter can be defined with the help of the Einstein equations, and the sought matter has the form of an anisotropic
perfect fluid \cite{Chopovsky:2014yxa} with the EoS parameter $\omega_0=-1$ in the external/our space and $\omega_1$ in the internal space, correspondingly. The parameter
$\omega_1$ satisfies the following relation \cite{Chopovsky:2014yxa}:
\be{1.3}
-\mathcal{C}=(1+\omega_1)\kappa\hat\varepsilon'\, ,
\ee
where $\hat\varepsilon'$ is the energy density of the background matter. The primes will denote values related to the background matter and its fluctuations.
Obviously, in the case of the Ricci-flat internal space $\mathcal{C}=0$ the background matter is absent: $\hat\varepsilon'=0$. Therefore, the energy-momentum
tensor (EMT) of the unperturbed background matter can be written as follows:
\be{1.4}
\hat T'^{M}_{N}
=\hat \ve '(\delta^M_\lambda\delta^\lambda_N-\om_1 \delta^M_l\delta^l_N)\, .
\ee
Taking into account the form of the background metric \rf{1.1}, it can be easily verified that this tensor satisfies the background conservation equation: $\hat
\nabla_M \hat T'^M_N=0$.

It is worth noting that choosing different values of $\omega_1$ (with fixed $\omega_0 = -1$), we can simulate different forms of matter. For example, $\omega_1=1$
corresponds to the monopole form-fields (the so called Freund-Rubin scheme of compactification \cite{Freund:1980xh,Accetta:1986vq,Gunther:2003zn}). Additionally, the condition
\be{1.5} \mathcal{C} \left[2 - d(1+\om_1)\right] >0\, , \ee
on the one hand, provides the stable compactification of the curved internal space and, on the other hand,  ensures the positiveness  of the radion mass squared
\cite{Chopovsky:2014yxa}. It is worth noting that we should also include into our model the multidimensional cosmological constant since in its absence the radion effective
potential has no minimum and the internal space is not stabilized \cite{Chopovsky:2014yxa}.


\section{Perturbed model. Conservation equation}

Now, we perturb the above background by a gravitating mass. We suppose that this mass is compact in three-dimensional (our) space. With respect to the internal
space, this mass can be either localized in some part of the space or it can be spread out over the entire volume. In what follows, we will demonstrate that the
Einstein equations together with the conservation equations require for the considered model the fulfillment of the latter condition. This gravitating mass
simulates an astrophysical object (e.g., our Sun). Since we know that the energy density inside the Sun is much bigger than the pressure, we assume the dust-like
EoS in the external (our) space: $p_0=0$. This is the usual assumption to calculate the PPN parameters in GR \cite{Landau}. However, in the internal space the EoS
reads $p_1=\Omega \epsilon$, where $\Omega$ can take arbitrary values. Therefore, the EMT of this gravitating body has the following form:
\be{2.1}
\tilde T^M_N\equiv\delta \tilde T^M_N=\epsilon \delta^M_0\delta^0_N-p_1 \,\delta^M_l\delta^l_N,\quad p_1=\Om\epsilon\, .
\ee
Hereafter, the tilde denotes values related to the gravitating mass.

This gravitating mass perturbs the background metric:
\be{2.2}
\hat g_{MN}\,\,\,\mapsto\,\,\,g_{MN}\approx\hat g_{MN}+ \delta \hat g_{MN}\equiv \hat g_{MN} + h_{MN}\, , \quad h^M_K\equiv \hat g^{ML}h_{LK}
\ee
and the background matter:
\be{2.3}
\hat T'^M_N\,\,\,\mapsto\,\,\,T'^M_N\approx\hat T'^M_N+\delta T'^M_N=
(\hat \ve '+\ve'_1)(\delta^M_\lambda\delta^\lambda_N-\om_1 \delta^M_l\delta^l_N)\, ,
\ee
where $\ve'_1\equiv\delta\ve'$ is supposed to be the perturbation of the background matter energy density. To get \rf{2.3}, we suppose that the perturbation does
not change the nature of the background matter. For example, if it was radiation before the perturbation, it remains radiation after that. In other words, the
perturbation does not change the EoS of the background matter.

Obviously, the total perturbed EMT $T^M_N=T'^M_N+\tilde T^M_N$ must satisfy the conservation equation $\nabla_M T^M_N=0$:
\be{2.4}
\nabla_M T^M_N
\approx \hat \nabla_M (\delta \tilde T^M_N+\delta T'^M_N)+(\delta\Gamma^M_{MS}\hat T^S_N-\delta \Gamma^{S}_{MN}\hat T^M_S)=0\, ,
\ee
where $\nabla_M$ ($\hat \nabla_M$) is the covariant derivative with respect to the perturbed (background) metric $g_{MN}$ ($\hat g_{MN}$) and we also used the
equation $\hat\nabla_M \hat T^M_N=0$. In Eq. \rf{2.4} (as well as throughout the paper) we keep only linear perturbations. For the covariant derivatives we get:
\begin{equation}
\label{2.5}
\begin{aligned}
\hat \nabla_M\delta \tilde T^M_N=
(\de_0 \epsilon)\delta^0_N-\Om \,(\de_l \epsilon)\delta^l_N\, ,\\
\hat \nabla_M\delta T'^M_N= (\de_\lambda \ve'_1)\delta^\lambda_N-\om_1(\de_l \ve'_1) \delta^l_N\, .
\end{aligned}
\end{equation}
Taking into account the relation \rf{1.3}, we also obtain
\be{2.6}
\delta\Gamma^M_{MS}\hat T^S_N-\delta \Gamma^{S}_{MN}\hat T^M_S
=-\cfrac{\mathcal{C}}{\kappa}\left[\delta\Gamma^m_{m\lambda}\delta^\lambda_N-\delta\Gamma^\mu_{\mu l}\delta^l_N\right]\, ,
\ee
where (see, e.g., Eq. (A.4) in \cite{Chopovsky:2013ala})
\be{2.7}
\delta \Gamma^m_{m\lambda}=\cfrac{1}{2}\,\de_\lambda h^l_l\, ,\quad \delta\Gamma^\mu_{\mu l}=
\cfrac{1}{2}\,\de_l h^\mu_\mu\, .
\ee
Therefore, the conservation Eq. \rf{2.4} for the different components reads as follows:
\ba{2.8}
&{}&\de_0\left(\epsilon+\ve'_1-\cfrac{\mathcal{C}}{2\kappa}\, h^l_l\,\right)=0\, ,\\
&{}&\label{2.9}\de_{\widetilde\nu}\left(\ve'_1-\cfrac{\mathcal{C}}{2\kappa}\, h^l_l\,\right)=0, \quad \widetilde\nu=1,2,3\\
&{}&\label{2.10}\de_n\left(\Om \epsilon+\om_1\ve'_1-\cfrac{\mathcal{C}}{2\kappa}\, h^\lambda_\lambda\,\right)=0, \,\, n=4,5,\ldots,4+d.
\ea
From \rf{2.9} we immediately find
\be{2.11} \ve'_1-\cfrac{\mathcal{C}}{2\kappa}\, h^l_l=f(y), \ee
where the constant of integration $f(y)$ is some function of the internal coordinates only. As we have emphasized above, the gravitating body is compact in the
external (our) space. Therefore, all perturbations should tend to zero at $r\equiv\sqrt{\sum_{\widetilde\mu}(x^{\widetilde\mu})^2}\longrightarrow \infty$. Hence,
$f(y)$ may only be zero:
\be{2.12}
\ve'_1=\cfrac{\mathcal{C}}{2\kappa}\, h^l_l\, .
\ee
It is worth noting that this condition is the direct consequent of the conservation law in the case of the compact gravitating masses. Then, Eqs. \rf{2.8} and
\rf{2.10} read
\be{2.13}
\de_0 \epsilon=0
\ee
and
\be{2.14}
\de_n\left[\Om \epsilon+\cfrac{\mathcal{C}}{2\kappa}\left(\om_1h^l_l-h^\lambda_\lambda\right)\right]=0\, .
\ee
Eq. \rf{2.13} shows that the gravitating mass should have static energy density. In the case of the Ricci-flat internal space $\mathcal{C}\equiv 0$ (then,
obviously, background matter is absent: $\hat\varepsilon'=\varepsilon'_1=0$), we obtain
\be{2.15}
\mathcal{C}=0\quad \mapsto \quad \Om\de_n\epsilon=0\, .
\ee
Therefore, in this case the gravitating mass is uniformly distributed over the internal space if $\Om\neq0$.

\section{Perturbed model. Einstein equation and gauge condition}

Now, we turn to the Einstein equation for the perturbed model. To simplify mathematical calculations, we consider for a while a special case when the background
internal space is a two-sphere of radius $a$:
\be{3.1}
\hat g_{mn}dy^m\otimes dy^n = -a^2(d\xi^2+\sin^2\xi d\eta^2)\, .
\ee
Here, $\mathcal{C}=-1/a^2$. This simplification does not affect the main results of our paper, and in final formulas we will restore the case of arbitrary
Einstein spaces. The background geometry is perturbed by the gravitating mass. In accordance with Eq. \rf{2.13}, it has static energy density distribution:
$\epsilon =\epsilon({\bf r}, \xi, \eta)$. Hence, perturbed geometry is also static. In order to find the perturbed metric coefficients $h_{MN}$, we must solve the
linearized field equation. It is worth noting that we did not assume any ansatz for the form of  $h_{MN}$. We will define it from the mutual analysis of the
Einstein and conservation equations. In the  case of the two-sphere internal space, the linearized Einstein equation reads \cite{Chopovsky:2011hp}:
\be{3.2} \cfrac{1}{\kappa}\,\delta R_{MN}= \delta \tilde T_{MN}+\delta T'_{MN}-\cfrac{1}{4}\left[\delta \tilde T+\delta T' \right]\hat g_{MN}-\cfrac{1}{4}\left(
\hat T'+2\La_6 \right)h_{MN}\, , \ee
where $\kappa$ is the multidimensional gravitational constant \cite{Chopovsky:2014yxa}, $\delta T=\delta T^L_L$ and $\Lambda_6$ is the multidimensional cosmological constant
which is fine tuned with the energy density of the background matter \cite{Chopovsky:2014yxa}: $\Lambda_6=\omega_1\hat\varepsilon'$. The linearized perturbations of the Ricci
tensor $\delta R_{MN}$ are (see, e.g., Eq. (A.5) in \cite{Chopovsky:2013ala}):
\be{3.3}
\delta R_{MN}=
\cfrac{1}{2}
\left[-\hat \nabla_L\hat \nabla^Lh_{MN}-\left(\hat R^L_{\,\,NPM}+\hat R^L_{\,\,MPN}\right)h^P_L+\hat R_{PM}h^P_N+\hat R_{PN}h^P_M\right]\, ,
\ee
where we have used the De Donder gauge{\footnote{We fix the gauge to remove non-physical degrees of freedom in $h_{MN}$ \cite{Landau}. The physical results should
not depend on the choice of the gauge.}} \cite{Landau}:
\be{3.4}
\hat \nabla_Lh^L_N-\cfrac{1}{2}\,\de_Nh^L_L=0\, .
\ee

The substitution of Eqs. \rf{2.1} and \rf{2.3} in the right hand side of \rf{3.2} results in the following system of equations:
\begin{align}
\label{3.5}
&\cfrac{1}{\kappa}\,\delta R_{00}=
\cfrac{1}{2}\left[\cfrac{3+2\Om}{2}\,\epsilon+\om_1\ve'_1\right]\, ,\\
\label{3.6}
&\cfrac{1}{\kappa}\,\delta R_{0\tilde\mu}=0\, ,\\
\label{3.7}
&\cfrac{1}{\kappa}\,\delta R_{\tilde\mu \tilde\nu }=
\cfrac{1}{2}\left[\cfrac{1-2\Om}{2}\,\epsilon-\om_1\ve'_1\right] \delta_{\tilde\mu\tilde\nu}, \\
\label{3.8}
&\cfrac{1}{\kappa}\,\delta R_{\mu n }=
-\cfrac{1}{\kappa a^2}\,h_{\mu n}\, ,\\
\label{3.9}
&\cfrac{1}{\kappa}\,\delta R_{mn}=\cfrac{1}{2}\left\{\left[-\cfrac{1+2\Om}{2}\,\epsilon
-(2+\om_1)\ve'_1\right]\hat g_{mn}-\cfrac{2}{\kappa a^2}\,h_{mn}\right\}\, ,
\end{align}
where in Eqs. \rf{3.8} and \rf{3.9} we have used the relation \rf{1.3} with $\mathcal{C}=-1/a^2$.

Let us calculate now the left hand side of Eq. \rf{3.2} where $\delta R_{MN}$ is defined by \rf{3.3}. Simple but lengthy calculations result in the following
expressions:
\begin{align}
\label{3.10}
&\delta R_{\mu\nu}=\cfrac{1}{2}\triangle_5 h_{\mu\nu}\, ,\\
\label{3.11}
&\delta R_{\mu 4}=
\cfrac{1}{2}\left[\triangle_5 h_{\mu 4}-\cfrac{1}{a^2\sin^2\xi}\, h_{\mu 4}-\cfrac{2\cos\xi}{a^2\sin^3\xi}\,\de_5h_{\mu5}\right],
\end{align}
\begin{align}
\label{3.12}
&\delta R_{\mu 5}=
\cfrac{1}{2}\left[ \triangle_5 h_{\mu 5}+\cfrac{2}{a^2}\, \cfrac{\cos\xi}{\sin\xi}\,\de_4 h_{\mu 5}+\cfrac{2\cos\xi}{a^2\sin\xi}\,\de_5h_{\mu 4} \right], 
\end{align}
\begin{align}
\label{3.13}
&\delta R_{44}=
\cfrac{1}{2}\left\{
\triangle_5 h_{44}-\cfrac{2}{a^2\sin^2\xi}\left[h_{44}-\cfrac{h_{55}}{\sin^2\xi}\right]
-\cfrac{4\cos\xi}{a^2\sin^3\xi}\,\de_5h_{45}\right\},
\end{align}
\begin{equation}
\begin{aligned}
\label{3.14}
\delta R_{55}=
\cfrac{1}{2}\left[\triangle_5 h_{55}+\cfrac{2\cos^2\xi}{a^2\sin^2\xi}\,h_{55}
-\cfrac{4\cos\xi}{a^2\sin\xi}\,\de_4 h_{55}+\cfrac{2}{a^2}\, h_{44}\right.\\
+\left.\cfrac{4\cos\xi}{a^2\sin\xi}\,\de_5h_{45}\right],
\end{aligned}
\end{equation}
\begin{equation}
\begin{aligned}
\label{3.15}
\delta R_{45}=
\cfrac{1}{2}\left\{ \triangle_5 h_{45}
+\cfrac{1}{a^2}
\left[-\cfrac{3}{\sin^2\xi}\,h_{45}
-2\cfrac{\cos\xi}{\sin\xi}\,\de_4h_{45}\right.\right.\\
-\left.\left.2\cfrac{\cos\xi}{\sin^3\xi}\,\de_5h_{55}+
2\cfrac{\cos\xi}{\sin\xi}\,\de_5h_{44}
\right] \right\}\, ,
\end{aligned}
\end{equation}
where $\de_4\equiv\partial/\partial \xi$, $\de_5\equiv\partial/\partial \eta$ and
\begin{equation}
\begin{aligned}
\label{3.16}
&\triangle_5 h_{\mu\nu}\equiv\triangle_3h_{\mu\nu}+\cfrac{1}{a^2}\,\triangle_2h_{\mu\nu}
\equiv
\sum_{\tilde\lambda=1}^3\de^2_{\tilde\lambda} h_{\mu \nu }
+\cfrac{1}{a^2}\,\de^2_4h_{\mu \nu }
\\&+\cfrac{1}{a^2\sin^2\xi}\left[\de_5^2h_{\mu \nu }
+\cos\xi\sin\xi\de_4h_{\mu \nu }\right]\, .
\end{aligned}
\end{equation}


Now, we want to make some preliminary conclusions. First, from Eqs. \rf{3.6}, \rf{3.7} and \rf{3.10} we obtain the following equation:
\be{3.17}
\triangle_5 h_{\mu\nu}=0\, ,\quad \mu\neq\nu .
\ee
Therefore, $h_{\mu\nu}=0$ for $\mu\neq\nu$. Second, from Eqs. \rf{3.8}, \rf{3.11} and \rf{3.12} we get the system of equations:
\ba{3.18}
\triangle_5 h_{\mu 4}-\cfrac{1}{a^2\sin^2\xi}\, h_{\mu 4}-\cfrac{2\cos\xi}{a^2\sin^3\xi}\,\de_5h_{\mu5}=-\cfrac{2}{a^2}\,h_{\mu 4}, \\
\label{3.19}
\triangle_5 h_{\mu 5}+\cfrac{2}{a^2}\, \cfrac{\cos\xi}{\sin\xi}\,\de_4 h_{\mu 5}+\cfrac{2\cos\xi}{a^2\sin\xi}\,\de_5h_{\mu 4}=-\cfrac{2}{a^2}\, h_{\mu 5}.
\ea
This system of equations shows that $h_{\mu 4}$ and $h_{\mu 5}$ serve as sources to each other but are decoupled from the EMT of the material sources. Therefore,
the perturbations of matter do not generate these fields and we should put:
\be{3.19a}
h_{\mu4}=h_{\mu 5}=0\, .
\ee

As we have already noted above, to  get expression \rf{3.3} we have used the gauge condition \rf{3.4}. Therefore we should check that our metric coefficients
satisfy this condition. Let us write this condition in components:
\begin{align}
\label{3.20}
\de_0\,\left(h_{00}+\sum_{\tilde \kappa}h_{\tilde \kappa\tilde \kappa}+\cfrac{1}{a^2}\, h_{44}+\cfrac{1}{a^2\sin^2\xi} \, h_{55} \right)=0,
\end{align}
\begin{align}
\label{3.21}
\de_{\tilde \nu} \,\left(h_{00}+2h_{\tilde\nu\tilde \nu }-\sum_{\tilde \kappa}h_{\tilde \kappa\tilde \kappa}-\cfrac{1}{a^2}\, h_{44}-\cfrac{1}{a^2\sin^2\xi}
\, h_{55} \right)=0,
\end{align}
\begin{equation}
\begin{aligned}
\label{3.22}
&\cfrac{1}{a^2\sin^2\xi}\,\de_5h_{45}+\cfrac{1}{a^2}\,\cfrac{\cos\xi}{\sin\xi}\,\left( h_{44}-\cfrac{1}{\sin^2\xi}\,h_{55}\right)\\
&+\cfrac{1}{2}\,\de_4
\left(h_{00}-\sum_{\tilde \kappa}h_{\tilde \kappa\tilde \kappa}+\cfrac{1}{a^2}\, h_{44}-\cfrac{1}{a^2\sin^2\xi} \, h_{55} \right)=0,
\end{aligned}
\end{equation}
\begin{equation}
\begin{aligned}
\label{3.23}
&\cfrac{1}{a^2}\left(\de_4h_{45}+\cfrac{\cos\xi}{\sin\xi}\,h_{45} \right) \\
&+\cfrac{1}{2}\,\de_5\,\left(h_{00}-\sum_{\tilde \kappa}h_{\tilde \kappa\tilde \kappa}-\cfrac{1}{a^2}\, h_{44}+\cfrac{1}{a^2\sin^2\xi} \, h_{55}  \right)=0.
\end{aligned}
\end{equation}
Eq. \rf{3.20} is satisfied identically for the static perturbations. From Eq. \rf{3.21} it follows that the expression in brackets is a function $\Phi(y)$ of the
internal coordinates only. However, for a compact source, the perturbations must decay at $|{\bf r}_3|\rightarrow\infty$, hence, we should put $\Phi(y)\equiv0$.
Therefore, the sum  in brackets is equal to zero.

Before presenting the full system of field equations, we introduce new notations in order to simplify the form of these equations:
\ba{3.24}
&{}&h_{00}=A, \quad h_{11}=B, \quad h_{22}=C, \quad h_{33}=D\, ,\nn\\
&{}&h_{44}=E, \quad h_{55}=F, \quad h_{45}=G\, .
\ea

The combinations \rf{3.10} with \rf{3.5} and \rf{3.7}  give, respectively,
\ba{3.25}
&{}&\triangle_5 A =\kappa\left[\cfrac{3+2\Om}{2}\,\epsilon+\om_1\ve'_1\right]\, ,\\
\label{3.26}
&{}&\triangle_5 B=\triangle_5C=\triangle_5D=\kappa\left[\cfrac{1-2\Om}{2}\,\epsilon-\om_1\ve'_1\right]\, ,
\ea
which result in the conclusion that
\be{3.27}
B=C=D\, .
\ee
The combinations \rf{3.13}, \rf{3.14} and \rf{3.15} with \rf{3.9} give, respectively,
\begin{equation}
\begin{aligned}
\label{3.28}
\triangle_5 E -\cfrac{2}{a^2\sin^2\xi}\left[E -\cfrac{F }{\sin^2\xi}\right]
-\cfrac{4\cos\xi}{a^2\sin^3\xi}\,\de_5G \\
=\left[\cfrac{1+2\Om}{2}\,\epsilon
+(2+\om_1)\ve'_1\right]\kappa a^2-\cfrac{2}{a^2}\,E \, ,\\
\end{aligned}
\end{equation}
\begin{equation}
\begin{aligned}
\label{3.29}
\triangle_5 F +\cfrac{2\cos^2\xi}{a^2\sin^2\xi}\,F
-\cfrac{4\cos\xi}{a^2\sin\xi}\,\de_4 F +\cfrac{2}{a^2}\, E
+\cfrac{4\cos\xi}{a^2\sin\xi}\,\de_5G\\
=\left[\cfrac{1+2\Om}{2}\,\epsilon
+(2+\om_1)\ve'_1\right]\kappa a^2\sin^2\xi-\cfrac{2}{a^2}\,F ,\\
\end{aligned}
\end{equation}
\begin{equation}
\begin{aligned}
\label{3.30}
\triangle_5 G
+\cfrac{1}{a^2}
\left[-\cfrac{3}{\sin^2\xi}\,G
-2\cfrac{\cos\xi}{\sin\xi}\,\de_4G \right. \\
-\left.2\cfrac{\cos\xi}{\sin^3\xi}\,\de_5F +
2\cfrac{\cos\xi}{\sin\xi}\,\de_5E
\right] =
-\cfrac{2}{a^2}\,G\, .
\end{aligned}
\end{equation}
The gauge conditions \rf{3.21}, \rf{3.22} and \rf{3.23} read, respectively:
\ba{3.31}
&{}&A -B-\cfrac{E}{a^2}\, -\cfrac{F}{a^2\sin^2\xi} \, =0\, ,
\ea
\begin{equation}
\begin{aligned}
\label{3.32}
\cfrac{1}{a^2\sin^2\xi}\,\de_5G +\cfrac{1}{a^2}\,\cfrac{\cos\xi}{\sin\xi}\,\left( E -\cfrac{1}{\sin^2\xi}\,F \right)\\
+\cfrac{1}{2}\,\de_4
\left(A-3B+\cfrac{E}{a^2}\, -\cfrac{F}{a^2\sin^2\xi}  \right)=0,\\
\end{aligned}
\end{equation}
\begin{equation}
\label{3.33}
\cfrac{1}{a^2}\left(\de_4G +\cfrac{\cos\xi}{\sin\xi}\,G  \right)
+\cfrac{1}{2}\,\de_5\,\left(A-3B-\cfrac{E}{a^2}\, +\cfrac{F}{a^2\sin^2\xi}  \right)=0.
\end{equation}
This system is closed by the conservation equations \rf{2.12} and \rf{2.14}:
\begin{equation}
\label{3.34}
\ve'_1=\cfrac{1}{2\kappa a^2}\, \left(\cfrac{E}{a^2}+\cfrac{F}{a^2\sin^2\xi}\right)\, ,
\end{equation}
\begin{equation}
\label{3.35} 
\de_n\left\{\Om \epsilon+\cfrac{1}{2\kappa a^2}\left[\om_1\left(\cfrac{E}{a^2}+\cfrac{F}{a^2\sin^2\xi}\right)+A-3B\right]\right\}=0, \quad n=4,5.
\end{equation}


\section{Weak-field limit of black strings and black branes}

Let us analyze now the system of equations \rf{3.25}-\rf{3.35}. It can be easily seen that the following relation takes place:
\be{4.1}
\triangle_5 \left(\cfrac{F}{\sin^2\xi}\right)=\cfrac{1}{\sin^2\xi}\,\triangle_5 F-\cfrac{4\cos\xi}{a^2\sin^3\xi}\,\de_4 F+2F\cfrac{1+\cos^2\xi}{a^2\sin^4\xi}\, .
\ee
Then, Eq. \rf{3.29} can be written (after dividing by $\sin^2\xi$) in the form
\ba{4.2}
&{}&\triangle_5\left(\cfrac{F}{\sin^2\xi}\right) +\cfrac{2}{a^2\sin^2\xi}\left[ E-\cfrac{F}{\sin^2\xi} \right]+\cfrac{4\cos\xi}{a^2\sin^3\xi}\,\de_5 G\nn \\
&{}&=\kappa a^2\left[\cfrac{1+2\Om}{2}\,\epsilon
+(2+\om_1)\ve'_1\right]-2\,\cfrac{F}{a^2\sin^2\xi}\, ,
\ea
which has the structure similar to \rf{3.28}. Subtracting now \rf{4.2} from \rf{3.28}, we obtain the equation:
\begin{equation}
\begin{aligned}
\label{4.3}
\triangle_5\left[E-\cfrac{F}{\sin^2\xi}\right] -\cfrac{4}{a^2\sin^2\xi}\left[E -\cfrac{F }{\sin^2\xi}\right]\\
+\cfrac{2}{a^2}\left[E -\cfrac{F }{\sin^2\xi}\right]
=\cfrac{8\cos\xi}{a^2\sin^3\xi}\,\de_5G\, .
\end{aligned}
\end{equation}
In its turn, Eq. \rf{3.30} can be written in the form:
\begin{equation}
\begin{aligned}
\label{4.4}
\triangle_5 G
-\cfrac{3}{a^2\sin^2\xi}\,G
-2\cfrac{\cos\xi}{a^2\sin\xi}\,\de_4G -\cfrac{2}{a^2}\,G \\
=-
2\cfrac{\cos\xi}{a^2\sin\xi}\,\de_5\left[E-\cfrac{1}{\sin^2\xi}\,F\right]\, .
\end{aligned}
\end{equation}
From Eqs. \rf{4.3} and \rf{4.4}, it follows that the functions $G$ and $E-F\sin^{-2}\xi$ serve as sources to each other, but are decoupled from the material
source. Hence, they are both zero:
\be{4.5}
G=0, \quad E=\cfrac{F}{\sin^2\xi}\, .
\ee
Therefore, the internal space is perturbed in the conformal way. Additionally, taking into account that $h_{\mu\nu}=0$ for $\mu\neq\nu$ and $h_{\mu n}=0$, we
arrive at the conclusion that {\it{the perturbed metric retains the block-diagonal form}}. In our previous papers the block-diagonal form of the perturbed metric
was accepted either as an ansatz \cite{Chopovsky:2014yxa} or as a consequence of the assumption of uniform smearing (over the internal space) of the gravitating mass
\cite{Chopovsky:2011hp,Chopovsky:2012qb,Eingorn:2012ip}. In the present article, we demonstrate for the considered model that this statement follows directly from the Einstein equation and
gauge condition without both of these assumptions.

Now, taking into account Eqs. \rf{4.5} and \rf{3.31}, we obtain that the perturbed background energy density \rf{3.34} satisfies the condition
\be{4.6}
\ve'_1=\cfrac{1}{\kappa a^2 }\,\cfrac{E}{a^2}=\cfrac{1}{2\kappa a^2 }\left(A-B\right)\, .
\ee
Therefore, Eqs. \rf{3.25} and \rf{3.26} read
\ba{4.7}
&{}&\triangle_5 A =\cfrac{3+2\Om}{2}\,\kappa\epsilon+\cfrac{\om_1}{2 a^2}\left(A-B\right)\, , \\
\label{4.8}
&{}&\triangle_5 B=\cfrac{1-2\Om}{2}\,\kappa\epsilon-\cfrac{\om_1}{2 a^2}\left(A-B\right)\, .
\ea
From this system we obtain
\ba{4.9}
&{}&\triangle_5(A+B)=2\kappa\epsilon\, ,\\
\label{4.10}
&{}&\triangle_5(A-B)-\cfrac{\om_1}{a^2}\,(A-B)=(1+2\Om)\kappa\epsilon\, .
\ea
For the given energy density distribution $\epsilon =\epsilon({\bf r}, \xi, \eta)$, it is straightforward to solve the Laplace and Helmholtz equations \rf{4.9}
and \rf{4.10} in order to find $A$ and $B$, and then $E=a^2(A-B)/2$, $F=E\sin^2\xi$ and, finally, $\varepsilon'\sim (A-B)$. We can generalize these equations to
the case of  $d$-dimensional Einstein internal spaces. Then, Eq. \rf{4.9} is not changed (except for the evident substitution: $\triangle_5 \to \triangle_{3+d}$)
but Eq. \rf{4.10} takes the form
\be{4.11}
\triangle_{3+d}(A-B)- \cfrac{2\mathcal{C}\left[2-d(1+\om_1)\right]}{d+2}\left(A-B\right)=\cfrac{2d\left(1+2\Omega\right)}{d+2}\kappa\epsilon\, .
\ee
We notice that if the internal space is a $d$-sphere then $\mathcal{C}=-(d-1)/a^2$. For the metric coefficients $E$ and $F$ we get, respectively: $E=(A-B)/(\hat
g^{44} d)$ and $F=E \hat g^{44}/\hat g^{55}$, and $\varepsilon'$ is still proportional to $A-B$.

The two remaining gauge equations \rf{3.32} and \rf{3.33} result in the following condition:
\be{4.12}
\de_n(A-3B)=0, \quad n=4,5\, ,
\ee
which in the general case of an arbitrary Einstein internal space is
\be{4.13}
\de_n\left[A-(1+d)B\right]=0, \quad n= 4,5,\ldots ,4+d\, .
\ee
Finally, taking into account Eq. \rf{4.12} (\rf{4.13} in the general case) as well as Eqs. \rf{4.5} and \rf{4.6}, the conservation equation \rf{3.34} reads now:
\be{4.14}
\de_n\left[\Om \epsilon -\cfrac{\mathcal{C}\om_1}{2\kappa }\left(A-B\right)\right]=0, \quad n=4,5,\ldots ,4+d.
\ee

Now, depending on values of $\Omega$ and $\mathcal{C}$ (e.g., zero or nonzero), we can construct different scenarios. However, in the present article, we want to
concentrate on the case when our multidimensional model satisfies the gravitational tests in the Solar system with the same accuracy as General Relativity (GR).
More precisely, we have in mind the parameterized post-Newtonian (PPN) parameter $\gamma$. It is well known \cite{Landau,Will} that in GR $\gamma =B/A=1$. It can
be easily seen from Eqs. \rf{4.7} and \rf{4.8} that the demand $A=B$ will result in the ratio (it does not matter whether $\mathcal{C}$ is equal to zero or not)
$B/A=(1-2\Om)/(3+2\Om)$ (in the general case, $B/A=(1-d\Om)/[1+d(1+\Om)]$). Therefore, the only way to have $A=B$ is to choose $\Om =-1/2$. As it follows from
\rf{4.14}, in this case the energy density $\epsilon$ is a function of coordinates of the external/our space: $\epsilon =\epsilon({\bf r})$, i.e. the gravitating
body is uniformly smeared over the internal space. Then, the perturbed metric coefficients $A$ and $B$ also depend only on the spatial coordinates of the external
space (see Eg. \rf{4.9}), and they satisfy the Poisson equation
\be{4.15}
\triangle_3 A=\triangle_3 B=\kappa\epsilon({\bf r})=\kappa_N \epsilon_3({\bf r})\, ,
\ee
where $\epsilon_3({\bf r}) = \epsilon({\bf r})/V_{\mathrm{int}}$, $V_{\mathrm{int}}$ is the volume of the internal space, $\kappa_N =
\kappa/V_{\mathrm{int}}\equiv 8\pi G_N/c^4$ and $G_N$ is the Newton's gravitational constant. In the weak-field limit $\epsilon_3({\bf r})\approx \rho_3({\bf
r})c^2$ where $\rho_3({\bf r})$ is the three-dimensional rest mass density of the gravitating body. In the case of a point-like gravitating mass $\rho_3({\bf
r})=m\delta({\bf r})$, and the perturbed metric coefficient $A$ is connected with the Newton's gravitational potential: $A=2\varphi_N/c^2$ where $\varphi_N=-G_N
m/r$.

On the other hand, if we first put $\Om =-1/2$, then we get from \rf{4.10} and \rf{4.11} that $A=B$, and that the gravitating body is smeared over the internal
space (see Eq. \rf{4.14}). It is well known that the value $\Om =-1/2$ corresponds to the black strings ($d=1 \to \mathcal{C}=0$) and the black branes ($d\geq 2$,
$\mathcal{C}$ can be both zero and nonzero). Therefore, for this value of $\Om$ the above equations describe the metric coefficients of the black strings/branes
in the weak-field limit. We notice that the exact solutions can be found, e.g., in \cite{Eingorn:2010wi,Chopovsky:2012qb,Eingorn:2011vu}.

We want to present arguments in favour of the  black strings/branes choice. First, these multidimensional astrophysical objects satisfy the gravitational tests
for the PPN parameter $\gamma$ at the same level of accuracy as GR. Second, they are preferable from the thermodynamical point of view. It is well known that
compact nonrelativistic astrophysical objects such as our Sun have the dust-like EoS since the pressure inside them is much less than the energy density. In our
paper \cite{Eingorn:2012ip} we have shown that in the case of multidimensional models the gravitating masses acquire effective relativistic pressure in the external
space. Certainly, such pressure contradicts the observations. The equality $\Om =-1/2$ (i.e. tension) is the only possibility to preserve the dust-like equation
of state in the external space. These two points are the astrophysical aspects of the black strings/branes. The third point is connected with the cosmological
aspect. We know that the stabilization of the compact internal space is a necessary condition to be in agreement with observations (e.g., it provides the absence
of the fifth force \cite{Eingorn:2012ng}). In our case such stabilization is achieved with the help of the background matter with the energy-momentum tensor \rf{1.4}
and condition \rf{1.5} \cite{Chopovsky:2014yxa}. Obviously, this stabilization should be preserved in the presence of the gravitating masses, which are the black strings/branes
in our case. These objects, gathering in galaxies and groups of galaxies, form the large scale structure of the Universe. At sufficiently large scales (more than
200-300 Mpc{\footnote{See also \cite{Li:2015yha,Eingorn:2015hza} for the recent discussion of the scale of homogeneity in the Gpc range.}}) the Universe looks isotropic and
homogeneous and matter in it can be considered in the form of a perfect fluid. In the realistic models this perfect fluid corresponds to the Cold Dark Matter with
the dust-like EoS.  It is remarkable that the only perfect fluid which, on the one hand, has the dust-like EoS in the external/our space and, on the other hand,
does not spoil the internal space stabilization, is the perfect fluid with the parameter of EoS in the internal space $\Om =-1/2$. That is, it corresponds to the
black strings/branes \cite{Eingorn:2010wi,Zhuk,book}. This is the cosmological aspect of the black strings/branes.


\section{Conclusion}

In the present paper we investigated viability of Kaluza-Klein models from astrophysical and cosmological points of view. In other words, we asked the question:
Do the gravitating bodies in the Kaluza-Klein model correspond to the observed picture of the world? For this purpose, we have considered a compact gravitating
body on the multidimensional background. We have assumed that such gravitating masses correspond to non-relativistic astrophysical objects such as our Sun.
Therefore, the gravitational field created by these masses is weak. For such weak-field approximation it is natural to assume, similarly to usual 4-D models
\cite{Landau}, that the background four-dimensional part of spacetime is flat, while the $d$-dimensional internal space is the Einstein space which can be either
Ricci-flat or curved. To have the curved internal spaces, we need to introduce background matter. Moreover, this matter under certain conditions stabilizes the
internal space. Phenomenologically, this matter is considered in the form of a perfect fluid with the proper EoS. The gravitating body perturbs both background
geometry and background matter. It has the dust-like EoS in the external space (as it takes place for the nonrelativistic astrophysical objects) but an arbitrary
EoS parameter $\Om$ in the internal space. We have supposed that perturbations do not change the nature of the background matter, i.e. the EoS are preserved after
perturbations. This is the only assumption for our model. In contrast to our previous papers (e.g., \cite{Chopovsky:2011hp,Chopovsky:2012qb,Eingorn:2012ip,Chopovsky:2014yxa}), we did not impose any
additional restrictions on the form of the perturbed metric as well as on the distribution of the energy density of the gravitating body. All conclusions have
been obtained from the general analysis of the Einstein equation, gauge condition and the conservation equation. This is the main difference from previous works.
For example, the block-diagonality of the perturbed metric naturally follows from the general analysis. The smearing of the gravitating mass over the internal
space also follows from these equations under certain conditions.

The general analysis has resulted in a number of equations and conditions for the perturbed metric coefficients and energy density. The different choices of $\Om$
can realize different scenarios. However, the case $\Om =-1/2$ , that is the case of the black strings/branes, has a number of advantages. First, multidimensional
astrophysical objects with such tension in the internal space satisfy the gravitational tests for the PPN parameter $\gamma$ at the same level of accuracy as
General Relativity. Second, they are preferable from the thermodynamical point of view \cite{Eingorn:2012ip}. Third, averaging over the Universe, they do not destroy the
stabilization of the internal space. Therefore, the black strings/branes are preferable from the astrophysical and cosmological points of view.


\section*{Acknowledgements}

\"{O}A acknowledges the support by the Distinguished Young Scientist Award BAGEP of the Science Academy. AZ  acknowledges financial support from the Scientific
and Technological Research Council of Turkey (TUBITAK) in the scheme of Fellowships for Visiting Scientists and Scientists on Sabbatical Leave (BIDEB 2221). AZ
also acknowledges the hospitality of \.{I}stanbul Technical University (ITU) where parts of this work were carried out. The authors are grateful to Maxim Eingorn
for stimulating discussions and valuable comments.


\end{document}